\documentclass[prc,twocolumn,epsfig,nofootinbib]{revtex4}
\usepackage{graphics}
\usepackage{epsfig}
\usepackage{amsfonts}
\usepackage{amsmath}
\usepackage{bm}

\usepackage{color}

\begin{document}
\title{Deformation-induced splitting of isoscalar E0 giant resonance:
\\
Skyrme random-phase-approximation analysis}
\author{
J. Kvasil$^1$, V.O. Nesterenko$^2$, A. Repko $^{1,3}$,  W. Kleinig $^{2}$,
and P.-G. Reinhard$^4$}
\affiliation{$^1$
Institute of Particle and Nuclear Physics, Charles University, CZ-18000,
Prague 8, Czech Republic}
\affiliation{$^2$
Laboratory of Theoretical Physics, Joint Institute for Nuclear
Research, Dubna, Moscow region, 141980, Russia}
\email{nester@theor.jinr.ru}
\affiliation{$^3$
Institute of Physics, Slovak Academy of Sciences,  84511, Bratislava, Slovakia}
\affiliation{$^4$
  Institut f\"ur Theoretische Physik II, Universit\"at
  Erlangen, D-91058, Erlangen, Germany}

\date{\today}

\begin{abstract}
The deformation-induced splitting of isoscalar giant monopole
resonance (ISGMR) is systematically analyzed in a wide range of masses
covering medium, rare-earth, actinide, and superheavy axial deformed
nuclei. The study is performed within the fully self-consistent
quasiparticle random-phase-approximation (QRPA) method based on the
Skyrme functional. Two Skyrme forces, one with a large (SV-bas) and
one with a small (SkP) nuclear incompressibility, are considered. The
calculations confirm earlier results that, due to the deformation-induced
E0-E2 coupling, the isoscalar E0 resonance attains a double-peak
structure and significant energy upshift. Our results are compared
with available analytic estimations. Unlike earlier studies, we get a
smaller energy difference between the lower and upper peaks and thus a
stronger E0-E2 coupling. This in turn results in more pumping of E0
strength into the lower peak and more pronounced splitting of ISGMR.
We also discuss widths of the peaks and their negligible
correlation with deformation.
\end{abstract}


\maketitle

\section{Introduction}

The isoscalar giant monopole resonance (ISGMR) is of fundamental
interest because it provides information on the nuclear
incompressibility \cite{Bl80}, see early \cite{Ha01} and recent
\cite{Colo08,Av13,Stone14} reviews. In deformed nuclei, the ISGMR
exhibits an additional remarkable feature: it couples strongly with
the $\rm{K}^{\pi}=0^+$ branch of the isoscalar giant quadrupole
resonance (ISGQR), leading to a double-peak structure (splitting) of
the ISGMR strength \cite{Ha01}. In prolate nuclei, the high-energy
peak constitutes the basic ISGMR while the low-energy peak is produced
by the deformation-induced coupling of monopole (E0) and quadrupole
(E2) strengths. The energy difference between the peaks is larger than
the widths of ISGMR and ISGQR \cite{Ha01}. In well deformed nuclei,
both peaks carry significant fractions of the monopole strength. As a
result, the deformation-induced splitting of ISGMR the can be observed
experimentally, as e.g. in $^{154}$Sm \cite{Yo04}.

The main features of the E0-E2 coupling and related splitting of the
ISGMR were investigated about thirty years ago, theoretically
\cite{Ki75,Ab80,Ja83,Ni85,Za78} and experimentally
\cite{Ki75,Bu80,Ga84}. Various models were used: bare Q-Q interaction
\cite{Ki75}, adiabatic cranking \cite{Ab80}, variational \cite{Ja83},
fluid-dynamical \cite{Ni85} and quasiparticle random-phase
approximation (QRPA) with an effective interaction
\cite{Za78}. The fluid-dynamical study
\cite{Ni85} was especially successful. Being self-consistent and based
on a simple Skyrme functional, it provided reasonable numerical
results and useful analytical estimates.

The interest on  ISGMR in deformed nuclei was revived by i) the appearance of
new experimental data, e.g. for Sm \cite{It03,Yo04}, Mo \cite{Yo13}
and Cd \cite{Pa12,Lui_Cd} isotopes, and ii) the progress of modern
self-consistent mean-field models (relativistic, Skyrme, Gogny)
\cite{Ben03,Vre05}.  In recent self-consistent QRPA calculations, the
ISGMR in deformed Mg \cite{Peru08,Lo10,Gupta_15,Kva15}, Si
\cite{Peru08}, Zr \cite{Yosh10} and Nd-Sm \cite{Yosh13,Kva_Is}
isotopes as well as in $^{238}$U \cite{Peru11} was explored. The
results of early studies \cite{Ab80,Ja83,Ni85,Za78} were generally
confirmed. The importance of self-consistency was corroborated by showing
that a self-consistent treatment leads to a narrowing the ISGMR
splitting and thus to a much better agreement with the
experiment. Moreover, the influence of neutron excess on the ISGMR
properties was investigated \cite{Yosh10} and the combined effect of the
nuclear incompressibility K$_{\infty}$ and isoscalar effective mass
$m^*_0/m$ on the ISGMR splitting was scrutinized \cite{Yosh13}.

In the present paper, we present a systematic Skyrme-RPA analysis of
the double-peak structure of the ISGMR in prolate axially deformed
nuclei. In extension of previous studies, we cover a wide mass region
involving medium (Cd), rare-earth (Nd, Sm, Dy, Er, Yb), actinide (U, No)
and superheavy (Fl) nuclei.  This allows to check the main
trends and analytical estimations \cite{Ja83,Ni85} for the ISGMR.  The
analysis goes up to the region of superheavy nuclei where the ISGMR
had not yet been inspected.

Besides this survey of deformation splitting, we briefly discuss some
contradictions in experimental data on ISGMR from RCNP (Research
Center for Nuclear Physics at Osaka University) \cite{Pa12,It03} and
TAMU (Texas A\&M University) \cite{Yo04}. Both groups use the
($\alpha,\alpha')$ reaction and multipole decomposition analysis
(MDA) but get substantially different results for the detailed
structure of the ISGMR. For example, in well-deformed $^{154}$Sm TAMU
data \cite{Yo04} demonstrate a distinctive two-peak structure of ISGMR
while RCNP data \cite{It03} give only one monopole peak. The
discrepancies between TAMU and RCNP data for ISGMR in spherical (Sn,
Sm, Pb) and deformed (Sm) nuclei were addressed already in our
previous studies \cite{Kva_Is,Kva_PS}. Here we continue the discussion
using RCNP \cite{Pa12} and TAMU \cite{Lui_Cd} results for Cd isotopes.
In \cite{Pa12,Lui_Cd}, these isotopes are treated as spherical, thus
no deformation splitting of ISGMR is assumed.  Following experimental data
\cite{bnl_exp} and our calculations, Cd isotopes have a
modest quadrupole deformation. Thus some weak
double-peak ISGMR structure might be expected here.

Our calculations are performed within the self-consistent
QRPA method \cite{Ri80} based on the Skyrme
energy-functional and specified for axially deformed nuclei
\cite{Repko}. In order to pin down the E0-E2 coupling, the
self-consistent {\it separable} random-phase approximation (SRPA)
model with Skyrme forces is used \cite{Ne02,Ne06}.  In both cases, the
pairing is treated within the Bardeen-Cooper-Schrieffer (BCS) scheme
\cite{Ben00}.

The paper is organized as follows. In Sec. 2, the calculation details
are outlined. In Sec. 3, the results for ISGMR strength functions are
discussed. It is demonstrated that just the E0-E2 coupling is responsible for
the double-peak structure of ISGMR. In Sec. 4, the trends for various characteristics
of ISGMR  are analyzed and compared with analytical estimations.
In Sec. 5, conclusions are given.

\section{Calculation scheme}

The calculations are performed within a two-dimensional (2D) QRPA
approach \cite{Repko}. The method is fully self-consistent because: i)
both the mean field and residual interaction are obtained from the
same Skyrme functional, ii) the residual interaction includes all the
terms of the initial Skyrme functional as well as the Coulomb direct
and exchange terms (the latter in the local-density approximation). Both
time-even and time-odd densities are taken into account.

The QRPA code employs a mesh in cylindrical coordinates.
The calculation box reaches three nuclear radii. The mesh size
is 0.4 fm for medium and rare-earth nuclei, 0.7 fm for U and No
and and 1.0 fm for superheavy Fl. The single-particle
spectrum implements all the levels from the bottom of the potential well
up to +30 MeV. Pairing with a contact $\delta$-force interaction
(also called as volume pairing) is treated at the BCS
level \cite{Ben00}. The pairing particle-particle channel is taken
into account in the residual interaction.

The ISGMR and ISGQR are computed in terms of E0 and E2 strength functions
\begin{equation}
S_0(E\lambda; E) = \sum_{\nu} |\:\langle \nu | \hat{M}(E\lambda) | 0
\:\rangle |^2 \: \xi_{\Delta}(E-E_\nu ) \label{4}
\end{equation}
where $\hat{M}(E0)=\sum_{i}^A (r^2 Y_{00})_i$ and $\hat{M}(E2)=\sum_{i}^A
(r^2 Y_{20})_i$ are isoscalar (T=0) transition operators, $|0\rangle$
is the ground state wave function, $|\nu\rangle$ and $E_{\nu}$ are
QRPA states and energies. The strength functions include a Lorentz
folding with
$\xi_{\Delta}(E-E_{\nu})=\Delta/(2\pi [(E-E_{\nu})^2+\Delta^2/4])$ and
a folding width $\Delta$.  The Lorentz function approximately
simulates smoothing effects beyond QRPA (coupling to complex
configurations and escape widths) and so allows comparison of
calculated and experimental strengths. In the present study, an
averaging of $\Delta$= 2 MeV is found optimal. The same folding width
was used in the previous studies of the ISGMR
\cite{Yosh13,Kva_Is,Kva_PS}.

\begin{table}
\begin{center}
\begin{tabular}{l|ccccc}
\hline
  & $K_{\infty}$ & $m^*_0/m$ & $J$ & $L$ & $\kappa_\mathrm{TRK}$\\
  & [MeV] & & [MeV]  & [MeV] & \\
\hline
SV-bas & 234 &  0.9 &   30 &  32 &  0.40 \\
SkP &    202 &  1.0 &   30 &  20 &  0.35 \\
\hline
\end{tabular}
\end{center}
\caption{\label{tab:NMP} The key parameters of symmetric nuclear
  matter (incompressibility $K_{\infty}$, effective mass $m^*_0
  /m$, symmetry energy $J$, slope of symmetry energy $L$, TRK sum-rule enhancement
  $\kappa_\mathrm{TRK}$) for the two Skyrme parameterizations used in
  this paper.}
\end{table}
Two Skyrme forces, SV-bas \cite{Kl09} and SkP,  with
$\delta$-force pairing \cite{SKPd} are used. Their key properties are
characterized by nuclear matter parameters given in Table
\ref{tab:NMP}. It is seen that these two forces essentially differ
by their incompressibilities.

The QRPA calculations employ a large configuration space with
particle-hole (two-quasiparticle) energies up to 70-75 MeV.
Depending on the nucleus,
the space involves 8500-9700 configurations with K=0.
The spurious mode lies below 2-3 MeV, i.e. safely beyond the
ISGMR structures located at 9-20 MeV. For
SV-bas, the monopole strength summed in the relevant
energy interval 9-45 MeV exhausts the energy weighted sum rule $
\rm{EWSR} =\hbar^2/(2 \pi m)\:A\: \langle r^2 \rangle_0 $
 by 100-105$\%$. A similar result is obtained for SkP.

At one place, the quasiparticle separable RPA model (SRPA)
\cite{Ne02,Ne06} is also used. SRPA exploits a self-consistent
factorization of the residual
interaction, which drastically reduces the computational expense
while keeping high accuracy of the calculations.  The method has been
successfully applied for description of ISGMR in spherical
\cite{Kva_PS} and deformed \cite{Kva_Is} nuclei.  Here we  employ
SRPA for analyzing purposes because this model can switch deliberately the
E0-E2 coupling \cite{Kva_Is} and so allows to scrutinize the deformation
effect in the splitting of the ISGMR.  We use in SRPA the same
calculational parameters (2D cylindrical mesh, size of the configuration
space, etc) as in QRPA.

The QRPA calculations are performed for 24 nuclei from
medium, rare-earth, actinide and superheavy regions.
Mainly well deformed nuclei are considered. The particular  isotopic chains
(Cd, Nd, Er, No, Fl) are involved. Some chains (Nd, Fl)
cover a transition from spherical to deformed nuclei.

For all nuclei, with exception of Cd isotopes, the equilibrium axial
quadrupole deformation $\beta$ is determined by minimization of the
total energy of the system. For soft Cd isotopes, our calculations
give shallow energy surfaces with very weak minima. Thus for the Cd chain
we use experimental deformation parameters \cite{bnl_exp}.

\section{Results and discussion}

We first consider the ISGMR strength functions
which allow a direct inspection of the deformation-induced splitting
of the resonance. The low-energy peak appearing due to E0-E2 coupling
and high-energy (main) peak tending to the ISGMR in the spherical limit
will be discussed.

\begin{figure}
\begin{center}
\includegraphics[width=20pc]{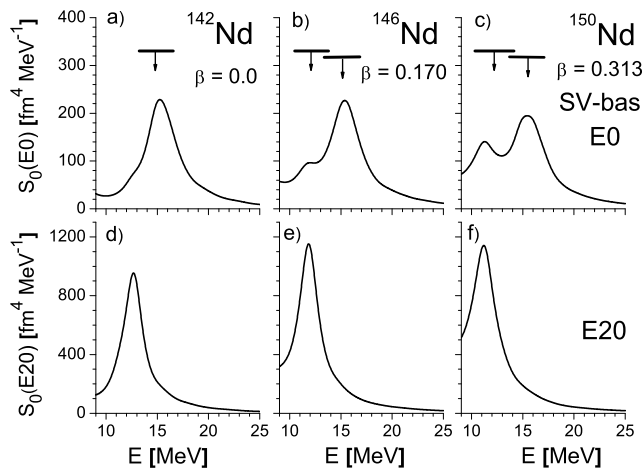}
\end{center}
\caption{\label{fig1} ISGMR (top panels)  and ISGQR(K=0) (bottom panels)
strength functions in $^{142,146,150}$Nd, calculated within QRPA.
The calculated deformation parameters are indicated
for each isotope. The experimental data (energy centroids and
widths) \cite{Ga84} are shown by arrows and horizontal bars.}
\end{figure}

In Fig.~\ref{fig1},
the double-peak structure of ISGMR is illustrated for the case of Nd
isotopes. Considered are spherical $^{142}$Nd, slightly deformed
$^{146}$Nd, and well deformed $^{150}$Nd. The low-energy peak in the
E0 strength grows with deformation. It is absent in spherical
$^{142}$Nd and significant in strongly deformed $^{150}$Nd where the
double-peak structure of the ISGMR becomes obvious.  The energy of the
low-energy peak precisely matches the position of K=0 branch of the
ISGQR (lower panels). This indicates that the low-energy peak in the
E0 strength is caused by the deformation-induced E0-E2 coupling
between ISGMR and ISGQR.

\begin{figure}
\begin{center}
\includegraphics[width=20pc]{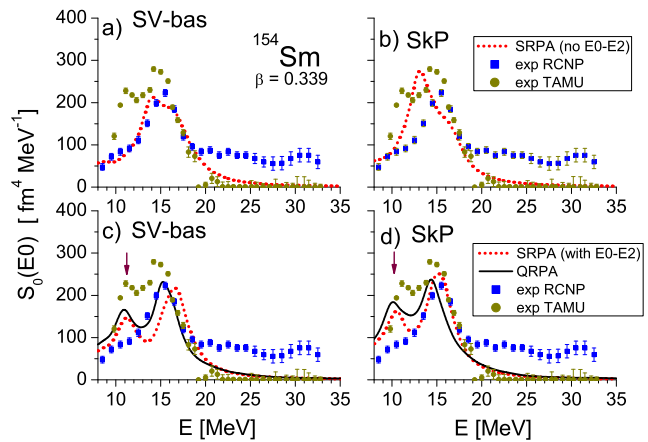}
\end{center}
\caption{\label{fig2}
Isoscalar E0 strength functions in deformed $^{154}$Sm
calculated within SRPA (red dotted lines) and QRPA (black solid lines)
with the forces SV-bas (left) and SkP (right). The SRPA calculations
are performed without (top panels) and with (bottom panels) the E0-E2 coupling.
For the comparison, the E0 data from  RCNP \protect\cite{It03} (blue filled squares)
and TAMU \protect\cite{Yo04} (green filled circles) experiments
are depicted.
In bottom panels, the calculated peak energies of $\lambda\mu$=20 branch
of ISGQR are marked by arrows.}
\end{figure}

The splitting of the ISGMR strength is further demonstrated in Fig.~\ref{fig2} for
well deformed $^{154}$Sm. This example is especially interesting since
for $^{154}$Sm two sets of experimental data, RCNP \cite{It03} and
TAMU \cite{Yo04}, are available. For convenience of the comparison, we
converted the TAMU data \cite{Yo04} given in fractions of the EWSR to
units fm$^4$ MeV$^{-1}$ used in RCNP \cite{It03}. In the upper panels,
the monopole strength from SRPA calculated without the E0-E2 coupling
is shown (note that, unlike QRPA, SRPA allows to suppress the E0-E2
coupling in deformed nuclei by skipping the quadrupole terms in the
input generators \cite{Ne06}).  In this case, the ISGMR, despite the
strong deformation, does not exhibit any distinctive double-peak
structure but instead comes  as one broad peak with some fine
structure. However, if we take into account the E0-E2 coupling (bottom
plots), then an additional low-energy peak appears and ISGMR attains a
double-peak structure. Moreover, the energy of the low-energy peak
coincides with the energy of ISGQR(K=0) resonance (marked in the plots
by arrows).  This takes place in both QRPA and SRPA calculations with
Skyrme parametrizations SV-bas and SkP. So Fig.~\ref{fig2}
once more proves that just E0-E2 coupling causes the double-peak
structure of ISGMR.

The lower panels of Fig.~\ref{fig2} also show that QRPA calculations
with SV-bas correctly reproduce the measured energies of
ISGMR peaks, while the calculations with SkP
underestimate them. So the incompressibility of SV-bas
K$_{\infty}$=234 MeV is more reasonable for $^{154}$Sm than the
SkP value K$_{\infty}$=202 MeV. It is worth mentioning that
the K$_{\infty}$=234 of SV-bas is in accordance with a good
reproduction of ISGMR in $^{208}$Pb together with the charge
formfactor in nuclear ground states \cite{Erl14a,Rei15c}.

\begin{figure*}
\begin{center}
\includegraphics[width=22pc]{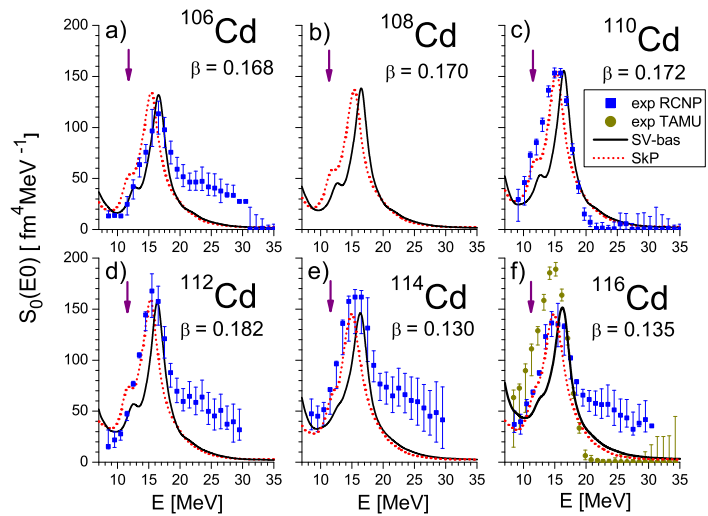}
\end{center}
\caption{\label{fig3}
Isoscalar E0 strength functions in Cd isotopes, calculated within QRPA
with the forces SV-bas (black solid line) and SkP (red dotted line).
For the comparison, the E0 data from  RCNP \protect\cite{Pa12} (blue filled squares)
and TAMU \protect\cite{Lui_Cd} (green filled circles, only for $^{116}$Cd) experiments
are depicted. The calculated peak energies of the quadrupole branch with K=0
are marked by arrows. The experimental deformation parameters $\beta$
\cite{bnl_exp} are indicated for each isotope.}
\end{figure*}

As seen from Fig.~\ref{fig2}, the RCNP \cite{It03} and TAMU
\cite{Yo04} data for ISGMR in $^{154}$Sm deviate from each
other. Unlike the TAMU data, the RCNP data i) give a strong tail of
the monopole strength above the ISGMR and ii) do not exhibit the ISGMR
splitting with its pronounced low-energy monopole peak, in spite of
the fact that $^{154}$Sm is strongly prolate. The discrepancy between
RCNP and TAMU data is surprising since both groups use  $(\alpha,
\alpha')$ reaction and exploit similar multipole decomposition
techniques to extract the monopole strength \cite{It03,Yo04}.  This
problem was briefly discussed in our previous studies of ISGMR in
spherical \cite{Kva_PS} and deformed nuclei \cite{Kva_Is}. As
mentioned in \cite{Kva_PS}, a possible reason for the discrepancy
could be that different incident energies of $\alpha$-particle beams
had been used in TAMU and RCNP experiments, 240 MeV \cite{Yo04} and
386 MeV \cite{It03} respectively.  Since $(\alpha,\alpha')$ is a
peripheral reaction, the TAMU and RCNP experiments may probe different
surface slices and thus experience different compression
responses. Indeed, the compression response depends on the nuclear
density which significantly varies in the surface region. The
TAMU/RCNP discrepancy, being yet unresolved, calls for additional
measurements and analysis.

Fig.~\ref{fig3} shows results for the chain of Cd isotopes.  Here
we use recent RCNP data \cite{Pa12} for $^{106,110,112,114,116}$Cd and TAMU
data \cite{Lui_Cd} for $^{116}$Cd. The data are obtained for
$\alpha$-particles with incident energies 400 MeV (RCNP) and 240 MeV
(TAMU). In the analysis of \cite{Pa12,Lui_Cd}, the open-shell
Cd isotopes are treated as spherical. However the experimental data
\cite{bnl_exp} give for these isotopes a modest axial quadrupole
deformation  which steadily grows from $^{106}$Cd
($\beta$=0.173) to $^{112}$Cd ($\beta$=0.186) and then sharply drops
toward $^{114}$Cd ($\beta$=0.130) and $^{116}$Cd ($\beta$=0.135). Note
that earlier experiments \cite{Raman} deliver similar deformations for
$^{110-112}$Cd but much larger for $^{114}$Cd ($\beta$=0.190) and
$^{116}$Cd ($\beta$=0.191). Anyway, unlike \cite{Pa12,Lui_Cd}, it is
more relevant to treat Cd isotopes as slightly deformed. As mentioned
above, our calculations demonstrate for these nuclei shallow energy
surfaces with too weak minima. So, in our analysis of ISGMR in
$^{106-116}$Cd, we use not calculated but experimental deformations
\cite{bnl_exp} given in Fig.~\ref{fig3}.

Fig.~\ref{fig3} shows that both SV-bas and SkP calculations
predict a slight low-energy peak related to the position of ISGQR(K=0)
mode.  The experimental strength also indicates a slight left
shoulder. However this shoulder is too small and vague to conclude
safely on a low-energy peak. So most probable nuclei with
$\beta<$0.2 cannot develop a measurable deformation splitting of
ISGMR.

Note that both SV-bas and SkP in general reproduce the energy E$_{\rm{ISGMR}}$
of the main monopole peak, though the difference between the predictions is
1-2 MeV. SV-bas well describes E$_{\rm{ISGMR}}$ in $^{106}$Cd but somewhat
overestimates it in $^{110-116}$Cd while SkP underestimates E$_{\rm{ISGMR}}$ in
$^{106}$Cd but performs better in $^{110-116}$Cd. So we meet here again a
well know problem that we do not get a simultaneous good description of ISGMR in
different nuclei with one and the same Skyrme force, see e.g. discussions
\cite{Av13,Colo08,Pa12,Kva_PS}. However, while in the previous studies
doubly-magic and open-shell nuclei were compared (with a noticeable compression
"softness" in open-shell patterns), here we see a different compression
"softness" already between open-shell nuclei.  Namely, ISGMR favors a lower
K$_{\infty}$ from SkP in Cd isotopes and larger K$_{\infty}$ from SV-bas in
$^{154}$Sm. Mind that the difference in effective mass $m^*_0/m$ does not affect
the ISGMR in spherical nuclei \cite{Kl09,Erl14a,Rei15c} so that $K_\infty$
remains as the only player in this case. But in deformed nuclei there is a
combined effect of K$_{\infty}$ and isoscalar effective mass $m^*_0/m$.  The
effective mass can affect ISGQR and thus the E0-E2 coupling. This in turn can
result in a noticeable shift of the main ISGMR peak. As discussed below, in
Cd isotopes we get a downshift of the main ISGMR peak (in contrast to well
deformed nuclei which demonstrate a noticeable upshift). Anyway our QRPA
analysis of ISGMR in Cd isotopes is still rather approximate.  A more
rigorous treatment should take into account a softness of these nuclei to deformation,
possible non-axiality, and coupling to complex configurations.

For $^{116}$Cd, Fig.~\ref{fig3}  once more exhibits big deviations between
RCNP \cite{Pa12} and TAMU \cite{Lui_Cd} experimental data. The deviations are
similar to those for $^{154}$Sm: as compared to TAMU,
RCNP gives a higher ISGMR energy and a strong strength tail above the resonance.
So, RCNP/TAMU deviations have a systematic character and are pertinent to
various mass regions. Presently this is a serious obstacle for further progress
in ISGMR studies.
\begin{figure}
\begin{center}
\includegraphics[width=20pc]{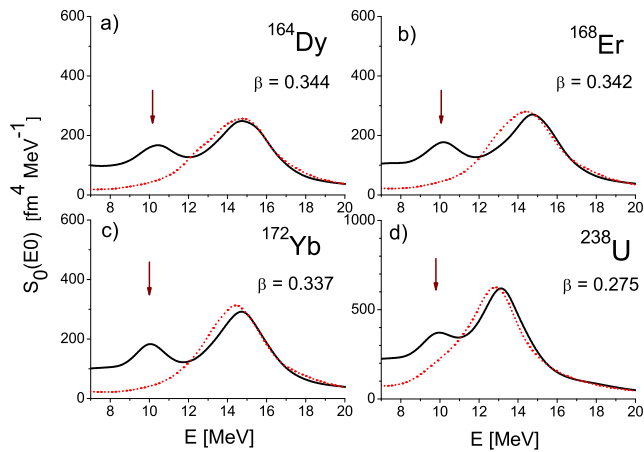}
\end{center}
\caption{\label{fig4}
The QRPA isoscalar E0 strength functions in deformed $^{164}$Dy,
$^{168}$Er, $^{172}$Yb and $^{238}$U, calculated with (black solid line)
and without (red dotted line) the equilibrium deformation (indicated in
the plots). The force SV-bas is used. The peak energies of
ISGQR(K=0) branch  are marked by arrows.}
\end{figure}

\begin{figure}
\begin{center}
\includegraphics[width=20pc]{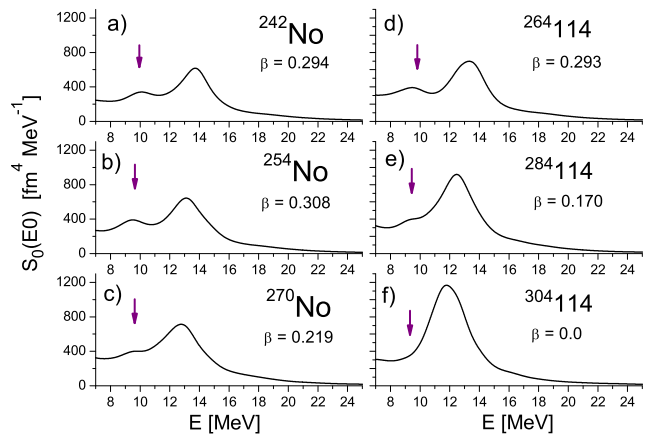}
\end{center}
\caption{\label{fig5}
The QRPA isoscalar E0 strength functions for isotopes
$^{242,254,270}$Nd (left)  and $^{264,284,304}$114 (right) calculated
at equilibrium deformations (indicated in the figure).
The force SV-bas is used. The peak energies of
ISGQR(K=0) branch  are marked by arrows.}
\end{figure}

In Figure~\ref{fig4}, the ISGMR in strongly deformed nuclei is
inspected.  Four typical rare-earth and actinide nuclei ($^{164}$Dy,
$^{168}$Er, $^{172}$Yb, $^{238}$U) are considered.  The E0 strength in
these nuclei demonstrates a clear two-peak structure and the
low-energy peak originates from the deformation-induced E0-E2 coupling.
In the spherical limit, only the main monopole peak exists.  Comparing
the strengths in the spherical and deformed cases, one can notice
that, in accordance to earlier studies \cite{Ha01,Ab80}, the E0-E2
coupling leads to some upshift of the main ISGMR peak.

In Figure~\ref{fig5},  SV-bas results for ISGMR are shown in
$^{242,254,270}$No and superheavy $^{264,284,304}$Fl (Z=114). The No chain
includes only deformed isotopes. The larger the isotope deformation,
the stronger the low-energy monopole peak. In all these isotopes, the
peak energy coincides with the energy of ISGQR(K=0) mode. The Fl chain
(Z=114) represents isotopes with zero (A=304), modest (A=284) and
large (A=264) deformation.  The low-energy peak is absent in spherical
$^{304}$114 but grows from $^{284}$114 to $^{264}$114. The peak energy
matches the ISGQR(K=0) energy thus confirming its origin from the
E0-E2 coupling. So superheavy nuclei confirm the physical
mechanism of ISGMR splitting pertinent to medium, rare-earth and actinide
nuclei.

\section{Trends}

Our calculations cover a wide mass region 106 $< A <$ 304 and various
deformations 0.17 $< \beta <$ 0.35. So they are suitable to analyze
the trends of the deformation splitting of ISGMR with mass number $A$
and quadrupole deformation $\beta$. We do this in terms of the following
key features of ISGMR and ISGQR in deformed nuclei:
\\
- the peak energy $E_{\rm{M}}$ of the high-energy (main) ISGMR bump
(tending to the principle ISGMR in the spherical limit),
\\
- the peak energy $E_{\rm{Q}}$ of the K=0 branch of ISGQR,
\\
- the deformation-induced splitting $\Delta{E}_M$ of the monopole strength
(the difference between peak energies of low- and high-energy ISGMR bumps),
\\
- the deformation-induced shift $dE_M$ of the high-energy ISGMR bump
(the difference between its peak energies computed with and without the deformation),
\\
- the fractions $S_1$ and $S_2$ of E0 strength of low- and high-energy ISGMR bumps,
\\
- the widths $w_1$ and $w_2$ of the low- and high-energy ISGMR bumps.
\\
The method of calculation  of the fractions and widths is described below
in discussion of Figs.~\ref{fig9} and~\ref{fig10}.

We use for the analysis the nuclei considered above and
additionally: $^{148}$Nd ($\beta$=0.22), $^{156}$Er ($\beta$=0.215),
$^{160}$Er ($\beta$=0.304), and $^{164}$Er ($\beta$=0.337).
Altogether the list of the nuclei covers
$^{106,108,110,112,114,116}$Cd, $^{142,146,148,150}$Nd, $^{154}$Sm,
$^{164}$Dy, $^{156,160,164,168}$Er, $^{172}$Yb, $^{238}$U,
$^{242,254,270}$No and $^{264,284,304}$Fl. The isotopic chains
for medium (Cd), rare-earth (Nd, Er), heavy (Nd), and superheavy
(Z=114) regions are considered.

\subsection{Model and empirical estimations}

The main characteristics of the impact of deformation on the ISGMR
were estimated about three decades ago in various models,
for a review see \cite{Ha01}. Between them, estimations within the cranking
model (CM) \cite{Ab80}, variational method (VM) \cite{Ja83}, and
fluid-dynamical model in the simple scaling approximation (SS)
\cite{Ni85} are most often used.  SS is
self-consistent and based on simplified Skyrme functional. Here we
mainly focus on VM and SS estimates as most detailed and robust
ones. Following \cite{Ab80,Ja83,Ni85}, we consider the estimates in
terms of the deformation parameter $\delta=0.946\beta$.
Note that the estimates
below concern only energies. We were not able to find any reliable
analytical estimates for E0 strengths.

In SS \cite{Ni85}, the estimates for the observables
of our interest read:
\begin{subequations}
\begin{align}
\label{A1}
E_{\rm{M}}&\approx E_{\rm{M}}^0 [1 -\frac{2}{9}\delta^2 +\frac{4}{9}\gamma_M\delta^2]
\\
&\approx E_{\rm{M}}^0 [1 -0.22\delta^2 +\underline{1.23\delta^2}]
\label{A1a}
\\
&\approx E_{\rm{M}}^0[1+1.01\delta^2] \; ,
\label{A2}
\\
& \gamma_{\rm{M}}=\frac{(E_{\rm{M}}^0)^2}{(E_{\rm{M}}^0)^2-(E_{\rm{Q}}^0)^2}
  =2.777
\end{align}
\end{subequations}
\begin{subequations}
\begin{align}
E_{\rm{Q}}
&\approx
E_{\rm{Q}}^0 [1-\frac{1}{3}\delta -\frac{1}{18}\delta^2-\frac{4}{9}\gamma_Q\delta^2]
\label{A3}
\\
&\approx E_{\rm{Q}}^0 [1-0.33\delta -0.06\delta^2-\underline{0.79\delta^2}]
\label{A4}
\\
 & \approx E_{\rm{Q}}^0 [1 - 0.33\delta - 0.85\delta^2]\; ,
 \label{A5}
\\
& \gamma_{\rm{Q}}=\frac{(E_{\rm{Q}}^0)^2}{(E_{\rm{M}}^0)^2-(E_{\rm{Q}}^0)^2}
 =1.777
 \end{align}
\end{subequations}
\begin{equation}
dE_{\rm{M}} \approx E_{\rm{M}}^0 1.01 \delta^2\; ,
\label{A6}
\end{equation}
\begin{subequations}
\begin{align}
\Delta E_{\rm{M}}
&\approx E_{\rm{M}}-E_{\rm{Q}}
\label{A7}
\\
&\approx E_{\rm{M}}^0[0.2 + 0.27\delta -0.18\delta^2 +\underline{1.87\delta^2}]
\label{A8}
\\
&\approx E_{\rm{M}}^0[0.2 + 0.27\delta + 1.69\delta^2]\; ,
\label{A9}
\end{align}
\end{subequations}
where
\begin{eqnarray}\label{E_ISGMR}
E^{0}_{\rm{M}}&\approx& 80 A^{-1/3} \rm{MeV},
\\
\label{E_ISGQR}
E^{0}_{\rm{Q}}&\approx& 64 A^{-1/3} \rm{MeV},
\end{eqnarray}
are empirical values for the ISGMR and ISGQR energies in spherical
nuclei \cite{Ha01}. The terms with $\gamma_M$ and $\gamma_Q$ in
Eqs. (\ref{A1}) and (\ref{A3}) arise due to the E0-E2 coupling (note
the relation $\gamma_{\rm{M}}/\gamma_{\rm{Q}}=(E^0_{\rm{M}}/E^0_{\rm{Q}})^2$).
In (\ref{A1a}), (\ref{A4}), and (\ref{A8}), these terms are
underlined to emphasize the impact of coupling. Note that values
$\gamma_{\rm{M}}$=2.777 and $\gamma_{\rm{Q}}$=1.777 are obtained from
empirical estimations (\ref{E_ISGMR}) and  (\ref{E_ISGQR}) (unlike \cite{Ni85}
where less realistic values $E^{0}_{\rm{M}}\approx 89 A^{-1/3} \rm{MeV}$ and
$E^{0}_{\rm{Q}}\approx 65 A^{-1/3} \rm{MeV}$ were implemented).
\begin{figure*}
\begin{center}
\includegraphics[width=28pc]{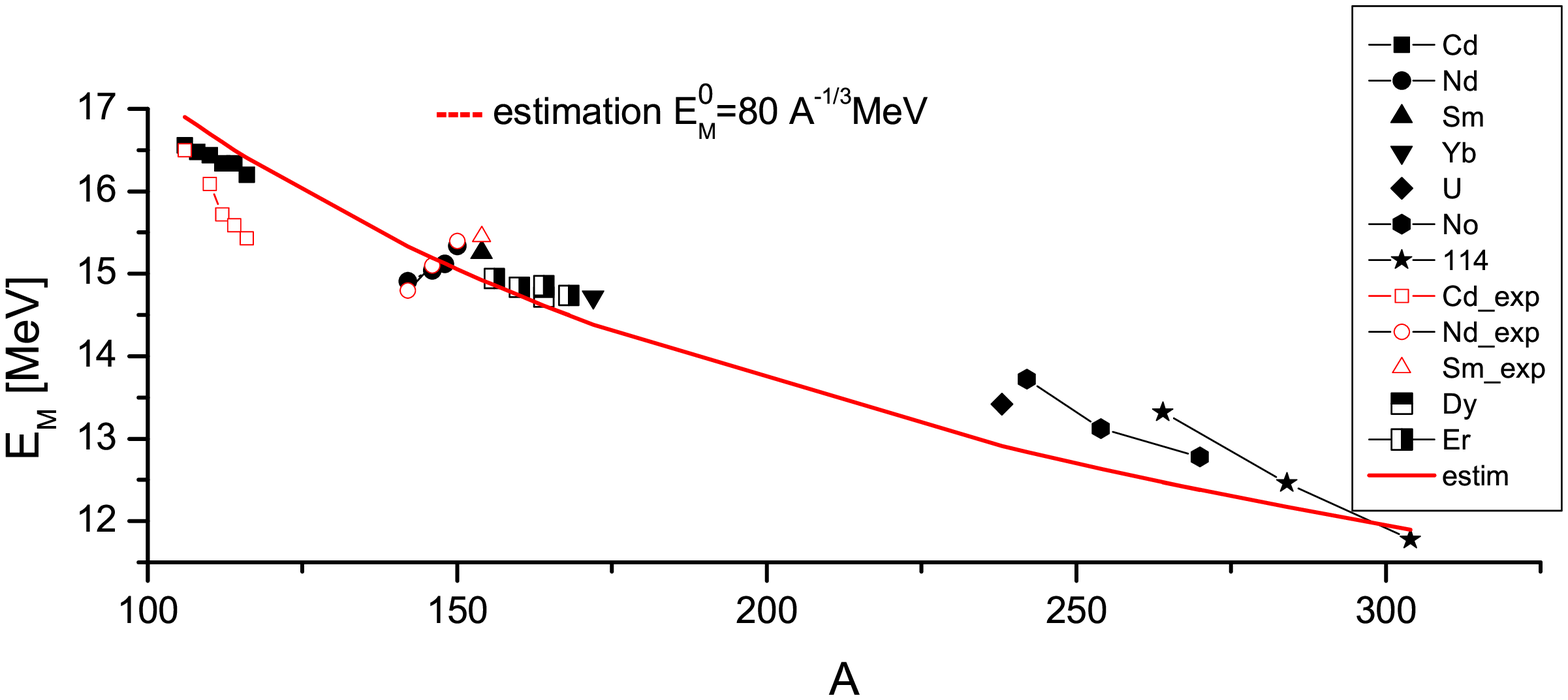}
\end{center}
\caption{\label{fig6} Dependence of the calculated ISGMR energies (black
filled symbols)  on the mass number A. For $^{106-116}$Cd, $^{142,146,148}$Nd
and $^{154}$Sm, the experimental values (red open symbols) are shown. The red line
gives the estimate $E_{\rm{M}}^0 \approx 80 A^{-1/3}$ MeV.}
\end{figure*}

The VM \cite{Ja83} provides
the estimates:
\begin{eqnarray}
E_{\rm{M}}&=&E_{\rm{M}}^0 (1 + 0.86\delta^2) ,
\label{A10}
\\
\label{A11}
E_{\rm{Q}}&=&E_{\rm{Q}}^0 (0.93 - 0.27\delta - 0.26\delta^2) ,
\\
dE_{\rm{M}}&=&E_{\rm{M}}^0 0.86 \delta^2 ,
\label{A12}
\\
\Delta E_{\rm{M}} &=&= E_{\rm{M}}^0 (0.26 + 0.22\delta + 1.1\delta^2) .
\label{A13}
\end{eqnarray}
For completeness, we quote also the recent empirical systematics
\cite{Scamps_PRC_14}
for the ISGQR(K=0) energy, obtained from self-consistent calculations
with Skyrme forces,
\begin{equation}
\label{Scamps}
E_{\rm{Q}}=E_{\rm{Q}}^0 (1 - 0.240 \delta - 0.672 \delta^2).
\end{equation}

Equations (\ref{A1})-(\ref{A13}) show that, despite some modest
differences in numerical coefficients, SS and VM lead to similar
important qualitative conclusions:
\\
i) Due to their large numerical coefficients, the $\delta^2$ -
corrections cannot be omitted.
For well deformed nuclei with $\delta \sim 0.3$,
these terms become of the same order of magnitude as the linear
$\delta$-terms, see (\ref{A5}), (\ref{A9}), (\ref{A11}), (\ref{A13}),
(\ref{Scamps}).  Moreover, in (\ref{A2}), (\ref{A6}), (\ref{A10}), and
(\ref{A12}), the terms linear in $\delta$ are absent at all and
deformation corrections are represented only by the quadratic terms.
\\
ii) The $\delta^2$-corrections take place even without E0-E2 coupling.
But then
their effect is small.  The E0-E2 coupling delivers additional large
$\delta^2$-terms which render the $\delta^2$-terms significant.
\\
iii) Just the E0-E2 coupling changes the sign and magnitude of the shift
$dE_{\rm{M}}$ as well as the sign and magnitude of the $\delta^2$-contribution
to the splitting $\Delta E_{\rm{M}}$.
\\
iv) As seen from (\ref{A8}), the E0-E2 coupling increases the
splitting $\Delta E_{\rm{M}}$. The same takes place  in CM and VM.
The result is natural  since interaction between
two levels always increases the energy distance between them \cite{Cast}.
\\
v) All deformation corrections in (\ref{A1})-(\ref{A13})
include $E^0_{\rm{M,Q}}$ and so exhibit the mass dependence $A^{-1/3}$.

For a large deformation $\delta$=0.3, the SS and VM give $dE_{\rm{M}} = 7.2
A^{-1/3}$MeV, $\Delta E_{\rm{M}} = 34.7 A^{-1/3}$MeV and $dE_{\rm{M}} = 6.2
A^{-1/3}$MeV, $\Delta E_{\rm{M}} = 34.0 A^{-1/3}$MeV, respectively. The
cranking model \cite{Ab80} gives $\Delta E_{\rm{M}} = 28.3 A^{-1/3}$MeV. So,
following these estimates, the deformation effect in ISGMR should be
strong. The upshift $dE_{\rm{M}}$ and deformation splitting $\Delta
E_{\rm{M}}$ can reach 10$\%$ and 35-45$\%$ of the resonance energy $E_{\rm{M}}^0$,
respectively. However, as noted in review \cite{Ha01},
the estimated splitting significantly exceeds the values from the experiment
for $^{154}$Sm and schematic QRPA calculations.
In this connection, it is interesting to
perform a systematic comparison of the estimates with the results of
self-consistent QRPA, which is just done below.

\subsection{QRPA trends}

In this section, we compare our QRPA results with the estimates given
above.  Following the notation of Sec. III, we deal here with the
deformation parameter $\beta$ instead of $\delta$.

In Fig.~\ref{fig6}, the dependence of the energy $E_M$ of the main
ISGMR peak on the mass number $A$ is exhibited and compared with the
estimate (\ref{E_ISGMR}). In general the calculated energies are well
aligned with the estimate, especially for spherical nuclei $^{142}$Nd
and $^{304}$Fl. However in deformed nuclei we get
$E_{\rm{M}}<E^{\rm{0}}_{\rm{M}}$ for $A<150$ and
$E_{\rm{M}}>E^{\rm{0}}_{\rm{M}}$ for $A>150$. These
deviations seem to be caused to a large extent by the deformation-induced shift
$dE_{\rm{M}}$ of the main ISGMR peak. Indeed, as shown below, $dE_{\rm{M}}$ is
negative for Cd isotopes and positive for deformed nuclei with $A>150$, which is in
accordance with the deviations.
In the isotopic chains for Nd and Er, the deformation and thus $dE_{\rm{M}}$ grow with A.
As a result, $E_{\rm{M}}$ in Er isotopes decreases with A slower than  $\sim A^{-1/3}$.
In Nd isotopes, the increase of $dE_{\rm{M}}$ with A even overrules the trend $\sim A^{-1/3}$
and, as a result, the resonance energy $E_{\rm{M}}$ grows with A,
both in experiment and our calculations.
Instead, in No and Fl isotopes the deformation and thus $dE_{\rm{M}}$ shrinks with A and so
these isotopic chains demonstrate even stronger decrease than $\sim A^{-1/3}$. So
the shift $dE_{\rm{M}}$ seems to play a noticeable role in determination of the energy
of the main ISGMR peak.
\begin{figure}
\begin{center}
\includegraphics[width=24pc]{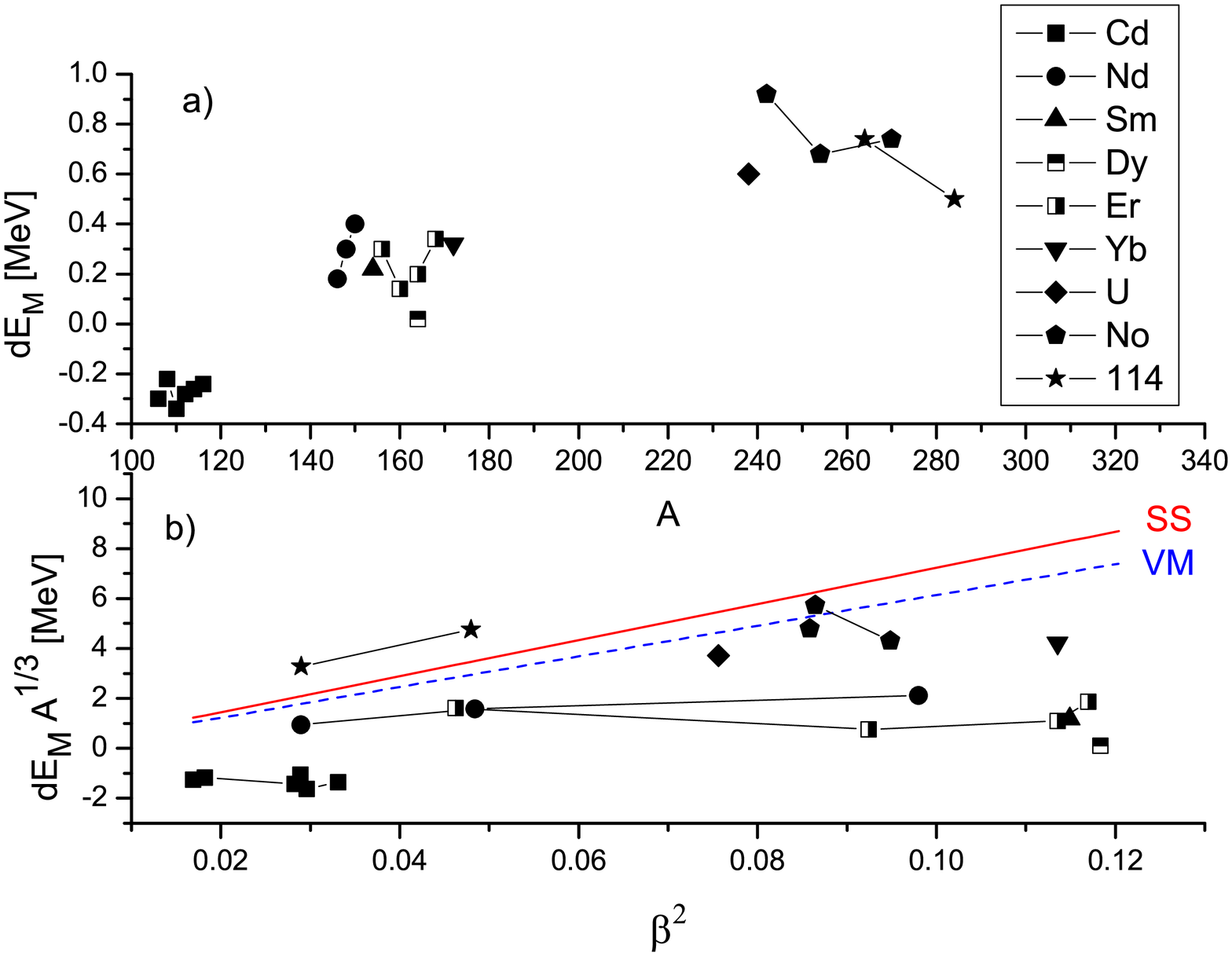}
\end{center}
\caption{\label{fig7}
a) Dependence of the calculated deformation shift $dE_{\rm{M}}$  on the mass number A.
(b) Dependence of $dE_{\rm{M}} A^{1/3}$ on the squared deformation parameter $\beta^2$.
 SS (red solid lines) and VM (blue dotted line)
 estimates are depicted.}
\end{figure}

\begin{figure}
\begin{center}
\includegraphics[width=20pc]{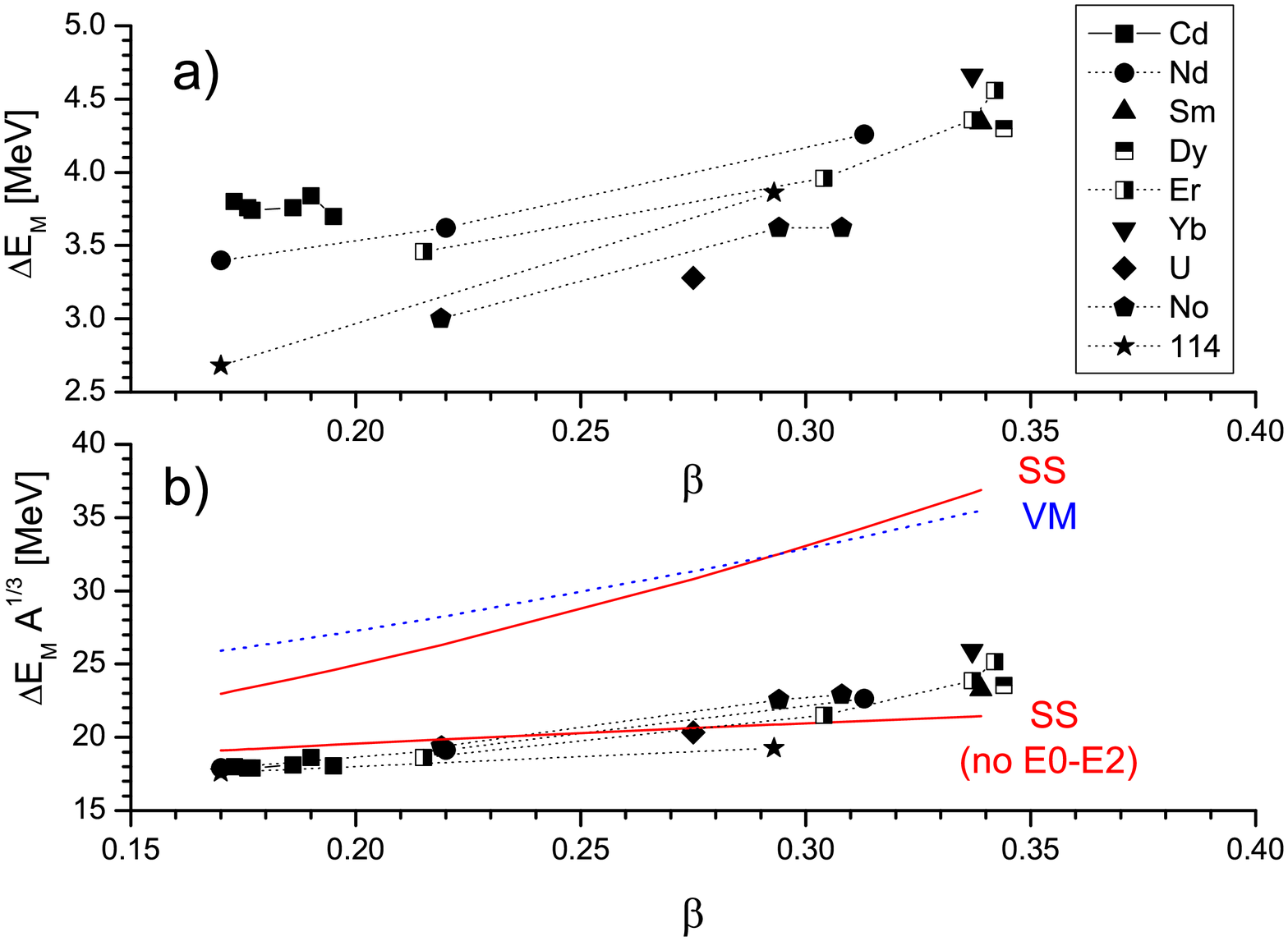}
\end{center}
\caption{\label{fig8}
(a) Dependence of the calculated deformation splitting
$\Delta E_{\rm{M}}$  (black filled symbols)  on the deformation parameter $\beta$.
 (b) The same but for $\Delta E_{\rm{M}} A^{1/3}$. SS estimates (\ref{A7})-(\ref{A9})
 and VM estimates (\ref{A13}) are depicted by red solid and blue dotted lines, respectively.
 For SS, the estimates with (upper line) and without
 (lower line) E0-E2 coupling are shown. See more details in the text.}
\end{figure}

The shift $dE_{\rm{M}}$ is inspected in detail in Fig.~\ref{fig7}.  The upper
panel shows that it varies between -0.4 and 1.0 MeV, depending on the
nucleus and its deformation. What is interesting, $dE_{\rm{M}}$ for Cd
isotopes is negative in contradiction with SS (\ref{A6}) and VM
(\ref{A12}) estimates. This signals that SS and VM estimates are not
robust for modest deformations $\beta<0.2$. With exception of Cd, all
other nuclei give positive $dE_{\rm{M}}$, i.e. an upshift. Its value varies
from 0-0.6 MeV in rare-earth nuclei to 0.5-1.0 MeV in actinides and
superheavy nuclei. So, in contradiction with the estimates, $dE_{\rm{M}}$
rather increases than decreases with $A$. There are exceptional cases: for
example, $^{164}$Dy demonstrates a negligible shift $dE_M$=0.02 MeV
despite a strong deformation $\beta$=0.344. So one may state that
SS/VM estimates for $dE_{\rm{M}}$ do not work at a quantitative level.

This is
additionally confirmed in the lower panel of Fig.~\ref{fig7} where the
values $dE_{\rm{M}} A^{1/3}$ are depicted.  Following the estimates, these
values should show a linear dependence on the squared
deformation parameter.  However, in the chains of Nd and Er isotopes,
$dE_{\rm{M}} A^{1/3}$ are about independent on $\beta^2$.
For other isotopes, a definite linear dependence on $\beta^2$ is also
not seen.  Besides the calculations give smaller values of $dE_{\rm{M}}$
than SS and VM estimates. Only heavy and superheavy
nuclei more or less match them.
Anyway in rare-earth, actinide and superheavy regions the calculated $dE_{\rm{M}}$ is
basically large. So this shift has to be taken into account in any
analysis of deformation effects in ISGMR.

The splitting $\Delta E_{\rm{M}}$ of the ISGMR is illustrated in
Fig.~\ref{fig8}. Both QRPA values as well as SS (\ref{A7})-(\ref{A9})
and VM  (\ref{A13}) estimates are given.  In QRPA,
$\Delta E_{\rm{M}}$ is determined as the difference between peak
energies of the low- and high-energy
ISGMR bumps (as is used in experimental data and
RPA calculations). Instead, in SS
and VM estimates, $\Delta E$
is defined as the difference between the ISGMR and ISGQR(K=0) energies.
The upper panel of Fig.~\ref{fig8} shows that
$\Delta{E}_{\rm{M}}$ lies in the range 2.5 - 5 MeV. Especially large
$\Delta{E}_{\rm{M}}$ is found in $^{174}$Yb ($\beta$=0.337). In $^{154}$Sm,
the computed splitting 4.34 MeV well agrees with the experimental value
4.12 MeV \cite{Yo04}.
The lower panel compares QRPA values of $\Delta E_{\rm{M}} \cdot A^{1/3}$ with SS
 and VM  estimates. The SS data are given
with and without E0-E2 coupling (i.e. with and without the underlined term in
 Eq. (\ref{A8}). The SS (with the coupling) and VM
estimates give similar results and both overshoot QRPA splitting by
20-35\%. This result is in accordance to the earlier finding \cite{Ha01}
that observed ISGMR splitting is only two-thirds of the estimates.
Only after removing E0-E2 coupling in SS this estimate comes close to the QRPA values.

Note that this finding should not be treated as an insignificance of E0-E2 coupling.
Instead all our results (spectacular two-bump structure of ISGMR caused by
E0-E2 coupling) testify that the coupling is strong and important.
Perhaps Fig.~\ref{fig8} rather indicates that SS overestimates the deformation effect,
which gives a false impression that the E0-E2 coupling is excessive.
This is partly confirmed by the comparison of $E_{\rm Q}$-values from
the empirical formula (\ref{Scamps}) and SS analytical
estimation (\ref{A5}): it is seen that (\ref{Scamps}) gives
much smaller deformation correction than (\ref{A5}).

\begin{figure}
\begin{center}
\includegraphics[width=20pc]{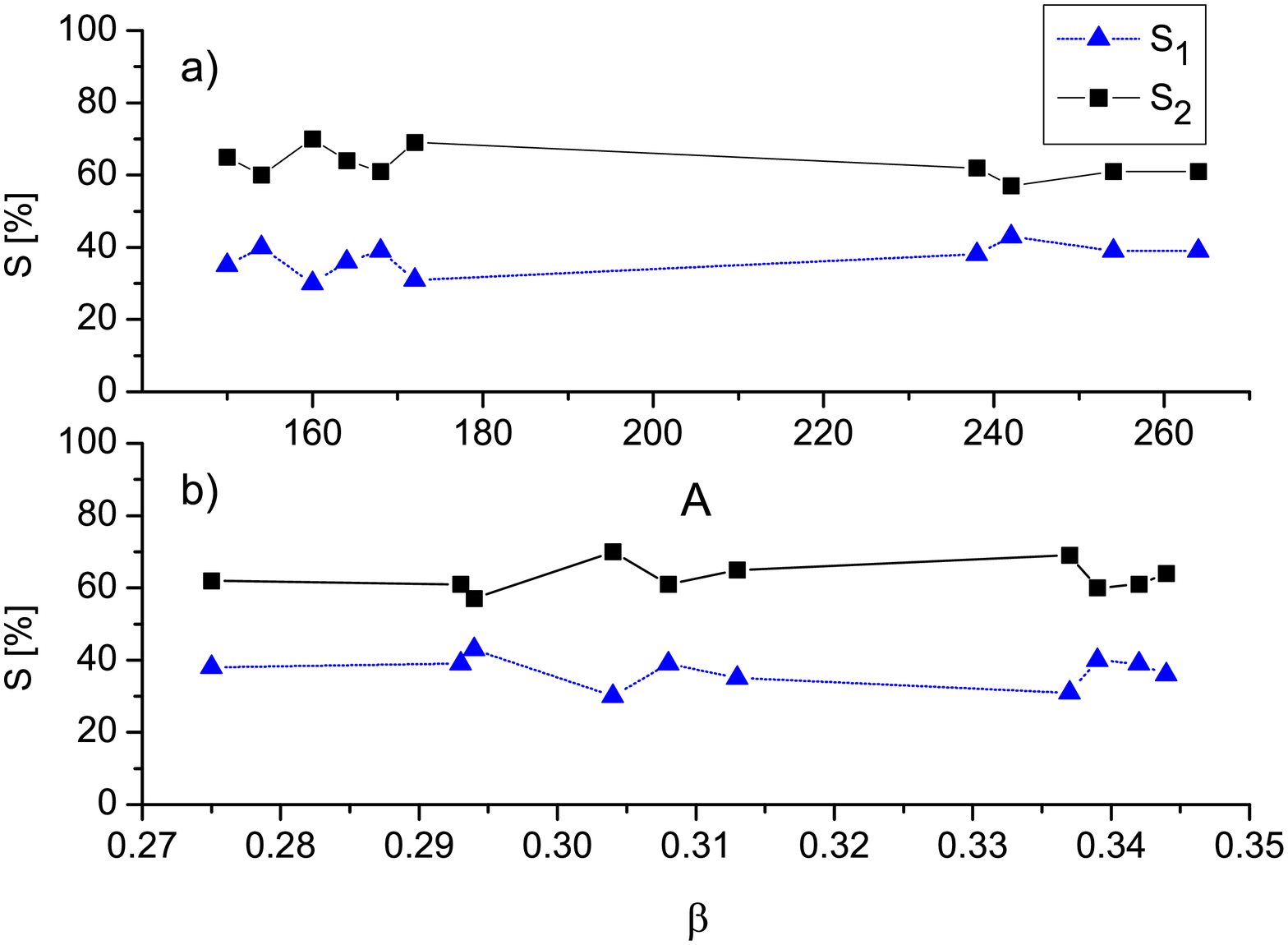}
\end{center}
\caption{\label{fig9}
Calculated fractions $S_1$ and $S_2$  (in $\%$)
of E0 strength in low- and high-energy branches of ISGMR in well deformed nuclei.
The dependence on the mass number (a) and deformation  (b)
is exhibited.}
\end{figure}
\begin{figure}
\begin{center}
\includegraphics[width=20pc]{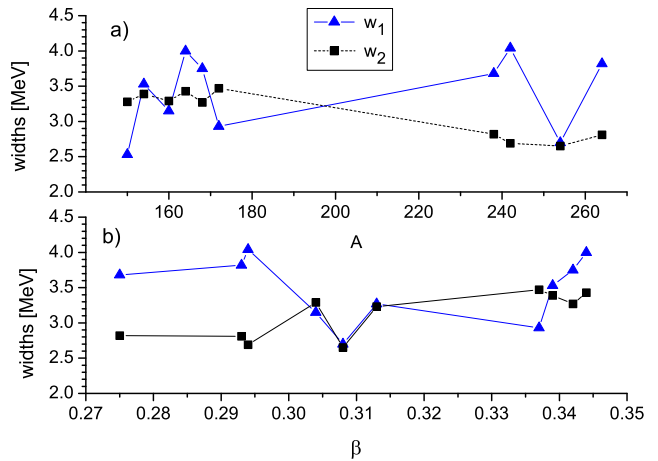}
\end{center}
\caption{\label{fig10}  Calculated widths $w_1$ and $w_1$ of lower
and upper ISGMR branches in well deformed nuclei.
The dependence on the mass number (a) and deformation  (b)
is exhibited.}
\end{figure}

Since, as compared to SS, QRPA produces a smaller energy difference between the lower and
upper ISQMR peaks, this should effectively lead to even stronger
coupling between them and to pumping more E0 strength into the lower
peak. This is demonstrated in  Fig.~\ref{fig9}, where
the calculated relative strengths $S_1$ and $S_2$ of the
lower and upper ISGMR branches are exhibited.  The strengths are obtained
by fitting the ISGMR
strength function by two Lorentzians at the energy interval 8-18 MeV, embracing
both ISGMR branches. To minimize ambiguities, only well deformed nuclei
with $\beta >$0.27 are considered. Because of
using a narrow interval $\beta$=0.27-0.35, the dependence of $S_1$ and
$S_2$ on $\beta$ is not so evident. However
Fig.~\ref{fig9} allows to demonstrate typical values of
$S_1$ and $S_2$ in different mass regions.
Remarkably large values of $S_1$ are obtained for
heavy and superheavy nuclei $^{284}$Fl $(\beta=0.170, S_1=28\%$),
$^{238}$U $(\beta=0.275, S_1=38\%$), and $^{242}$No $(\beta=0.294,
S_1=43\%$). So heavy and superheavy nuclei are promising samples to
observe the E0-E2 coupling. Besides that, large values of $S_1 \ge
35\%$ are obtained in well-deformed rare-earth nuclei $^{150}$Nd
$(\beta=0.313, S_1=35\%$), $^{154}$Sm $(\beta=0.339, S_1=40\%$,
as compared to the experimental value 32$\pm$2\% \cite{Yo04}),
$^{164}$Dy $(\beta=0.344, S_1=36\%$), and $^{168}$Er $(\beta=0.342,
S_1=39\%$). In general our calculations for well deformed nuclei give
larger $S_1 \approx 35-40\%$ as compared with $S_1 \approx 20-25\%$ in
early estimates \cite{Ab80,Ja83,Ni85}.  As mentioned above, this can be
caused by a smaller energy distance between the lower and upper ISGMR
peaks in our QRPA calculations. At the same time, our results are in
reasonable agreement with available experimental data which in
average give a higher monopole peak twice stronger than the lower one
\cite{Ha01}.

Figure~\ref{fig10} shows the calculated widths $w_1$ and $w_2$ of the low- and
high-energy branches of ISGMR. The widths are also obtained by fitting
of the ISGMR strength function by two Lorentzians.
And again only nuclei with a large
deformation $\beta > $ 0.27 are considered.  Fig.~\ref{fig10}
shows that both widths vary in
the range 2.5 - 4 MeV, which is in accordance to the summary in
\cite{Ha01}.  In average, $w_1$ and $w_2$ are of the same order but
various ratios $w_1 >, \approx, < w_2$ are realized. Following the
bottom panel, there is no definite dependence of the widths on
the nuclear deformation.  Instead, they fluctuate from nucleus to
nucleus and seem to be mainly determined by the actual level structure
(spectral fragmentation). The absence of definite dependence on
$\beta$ may be explained by two opposite trends caused by the
deformation. On the one hand, the deformation should increase the
width of the main ISGMR branch. But, on the other hand, the larger the
deformation, the more E0 strength is transferred to the low-energy
branch, which effectively decreases the width of the main
resonance. Note that different ratios $w_1/w_2$ were also found in the
previous study \cite{Yosh13} for Nd-Sm isotopes.

\section{Conclusions}

The deformation-induced splitting of isoscalar giant monopole
resonance (ISGMR) was systematically investigated for a variety of
axially deformed nuclei in a wide mass region $106 \le A \le 304$,
including medium, rare-earth, actinide and superheavy nuclei.
Altogether 24 nuclei were involved. The analysis was carried out in
the framework of the self-consistent quasiparticle
random-phase-approximation (QRPA) method based on the Skyrme
functional \cite{Repko}. Some auxiliary calculations were performed
with the separable self-consistent QRPA \cite{Ne02,Ne06}.  The Skyrme
force SV-bas \cite{Kl09} with incompressibility K$_{\infty}$=234 MeV
and isoscalar effective mass $m_0^*/m$=0.9 was mainly used,
complemented by the results from the Skyrme force SkP$^\delta$ with lower
K$_{\infty}$=202 MeV and somewhat higher $m_0^*/m$=1.  To the best of
our knowledge, this is the first
systematic self-consistent study of the deformation-induced splitting
of ISGMR, covering nuclei from different mass regions.

The calculations confirmed earlier result \cite{Ha01} that
deformation-induced coupling between monopole and quadrupole
excitations leads to the splitting of ISGMR into low- and high-energy
branches. The energy of the lower branch coincides with the energy of
K=0 part of isoscalar giant quadrupole resonance (ISGQR).  This effect
was found in all considered deformed nuclei, including superheavy
ones.

Following our analysis, the splitting of ISGMR cannot be easily
discriminated in nuclei with modest deformation $\beta < 0.2$. In
particular, it can hardly be observed in soft Cd isotopes.  At the
same time, in rare-earth, actinide and superheavy mass regions there
are many well deformed ($\beta > 0.3$) nuclei where the splitting is
strong enough to be observed experimentally. Our calculations well
reproduce experimental distributions of the monopole strength in $^{154}$Sm
\cite{Yo04} and Cd isotopes \cite{Pa12}. However it should be noted
that available
experimental data on ISGMR in well deformed nuclei are still sparse
and even contradicting.  For example, in well deformed $^{154}$Sm a
clear ISGMR splitting takes place in the TAMU experiment \cite{Yo04}
but not in the data from RCNP \cite{It03}. Our QRPA results agree well
with the TAMU data.  Anyway, there is a need in further high-accuracy
experimental studies. Moreover these studies should cover more of the
well deformed nuclei.

Since our investigation involves nuclei from different mass regions, it allows
to check various systematics, e.g. the trends with the mass number $A$
and deformation $\beta$. In this connection, we analyzed and tested
early estimates of different characteristics of deformation splitting
\cite{Ja83,Ni85}. The qualitative features following from the
estimates were generally confirmed. At the same time essential
differences at the quantitative level were found. Altogether, our QRPA
analysis has revealed the following peculiarities of ISGMR splitting:

i) The E0-E2 coupling leads to a noticeable (from -0.4 to 1.0 MeV)
energy shift $dE_{\rm{M}}$ of the main (upper) ISGMR peak. In
particular for isotopic chains,
this shift can noticeably affect the dependence of the ISGMR energy
on mass number. The effect is strong and
should be taken into account in any analysis of ISGMR in deformed
nuclei.  Unlike the estimates \cite{Ja83,Ni85}, the shift $dE_{\rm{M}}$ is
negative in Cd isotopes.  In other nuclei, it is positive but not
proportional to the squared deformation parameter $\beta^2$ as was
predicted \cite{Ja83,Ni85}.

ii) The QRPA result for the energy splitting $\Delta E_{\rm{M}}$ of ISGMR is by 20-30 $\%$
smaller than estimated in simple models \cite{Ja83,Ni85}, which is
in accordance with earlier conclusions \cite{Ha01}.  This has an important
consequence: smaller energy distance between the lower and upper ISGMR
peaks results in stronger E0-E2 coupling and thus in more pumping of
E0 strength to the lower peak.

iii) As a result, QRPA predicts in well deformed nuclei a large
fraction of E0 strength in the lower peak.  This fraction can reach
30-40$\%$ as compared to $\sim 30\%$ in earlier studies
\cite{Ha01} and recent experiment for $^{154}$Sm \cite{Yo04}.
The large strength of the lower peak opens promising
perspectives for experimental observation of the splitting of ISGMR strength.

iv) In accordance to recent QRPA study of ISGMR in Nd and Sm isotopes
\cite{Yosh13}, our analysis shows that widths $w_1$ and $w_2$ of the lower
and upper ISGMR peaks are generally of the same order of magnitude
(2.5-4 MeV), though they may exhibit different ratios, depending on
the nucleus. It is remarkable that both widths do not demonstrate a
definite dependence on the deformation $\beta$. This can be explained,
at least for the upper peak, by two opposite trends caused by the
deformation. On the one hand, the deformation should increase the
width of the main upper ISGMR peak (deformation spread). But,
on the other hand, the larger
the deformation, the more E0 strength is transferred to the lower
peak, which effectively decreases the width of the main peak.

Finally note that the deformation-induced coupling of the monopole
and quadrupole giant resonances is the only example of a strong
and measurable coupling between giant resonances.  This coupling
is of crucial importance when using the ISGMR in deformed nuclei for the
exploration of the nuclear incompressibility. Further, the E0-E2 coupling
can be useful for a combined investigation of nuclear incompressibility and
isoscalar effective mass \cite{Yosh13}. The lower peak of ISGMR could be a
useful indicator of the K=0 branch of ISGQR.

The study of the deformation-induced ISGMR splitting in
self-consistent models still leaves many open questions. This subject
needs both a strong theoretical effort and new high-accuracy
experiments for a variety of deformed nuclei.

\section*{Acknowledgments}

The work was partly supported by the DFG grant RE 322/14-1,
Heisenberg-Landau (Germany-BLTP JINR), and Votruba-Blokhincev
(Czech Republic-BLTP JINR) grants. P.-G.R. is grateful for the
BMBF support under Contracts No, 06 DD 9052D and 06 ER 9063. The
support of the Czech Science Foundation (P203-13-07117S) is
appreciated. A.R. is grateful for the support by the
Slovak Research and Development Agency
under the contract No. APVV-15-0225.


\begin{thebibliography}{99}
\bibitem{Bl80} 
 J. Blaizot,  Phys. Rep. {\bf 64}, 171 (1980).
\bibitem{Ha01} 
  M.N. Harakeh and A. van der Woude, {\it Giant Resonances}
(Oxford: Clarendon, 2001), Chapter 5.
\bibitem{Colo08} 
  G. Col\`o,
  Phys. Part. Nucl. {\bf 39}, 286 (2008).
\bibitem{Av13} 
 P. Avogadro and C.A. Bertulani,
 Phys. Rev. {\bf C88}, 044319 (2013).
\bibitem{Stone14} 
J.R. Stone, N.J. Stone, and S.A. Moszkowski,
    Phys. Rev. {\bf C89}, 044316 (2014).
 \bibitem{Yo04} 
 D.H. Youngblood, Y.W. Lui, H.L. Clark, B. John, Y. Tokimoto, and X. Chen,
 Phys. Rev. {\bf C69}, 034315 (2004).
\bibitem{Ki75} 
 T. Kishimoto, Phys. Rev. Lett. {\bf 35}, 552 (1975).
\bibitem{Ab80} 
 Y. Abgrall, B. Morand, E. Caurier, and N. Grammaticos, Nucl.
 Phys. {\bf A346}, 431 (1980).
 \bibitem{Ja83} 
   S. Jang, Nucl. Phys. {\bf A401}, 303 (1983).
 \bibitem{Ni85} 
   S. Nishizaki and K. Ando, Prog. Theor. Phys.{\bf  73}, 889  (1985).
\bibitem{Za78} 
 D. Zawischa, J. Speth, D. and Pal, Nucl. Phys. {\bf A311}, 445 (1978).
 \bibitem{Bu80} 
  M. Buenerd, D. Lebrun, P. Martin, P. de Saintignon, and C. Perrin,
  Phys. Rev. Lett. {\bf 45}, 1667 (1980).
 \bibitem{Ga84} 
   U. Garg, P. Bogucki, J.D. Bronson, Y.W. Lui, and D.H. Youngblood,
   Phys. Rev. {\bf C29}, 93 (1984).
\bibitem{It03} 
 M. Itoh, H. Sakaguchi, M. Uchida, T. Ishikawa, T. Kawabata, T. Murakami,
 H. Takeda, T. Taki, S. Terashima, N. Tsukahara, Y. Yasuda, M. Yosoi,
 U. Garg, M. Hedden, B. Kharraja, M. Koss, B.K. Nayak, S. Zhu, H. Fujimura,
 M. Fujiwara, K. Hara, H.P. Yoshida, H. Akimune, M.N. Harakeh, and M. Volkerts,
 Phys. Rev. {\bf C68}, 064602 (2003).
\bibitem{Yo13} 
 D.H. Youngblood, Y.W. Lui, Krishichayan, J. Button, M.R. Anders,
 M.L. Gorelik, M.H. Urin, and S. Shlomo,
  Phys. Rev. {\bf C88}, 021301(R) (2013).
\bibitem{Pa12} 
 D. Patel, et al, Phys. Lett. {\bf B718}, 447 (2012).
 \bibitem{Lui_Cd} 
 Y.W. Lui, D.H. Youngblood, Y. Tokimoto, H.L. Clark, and B. John,
 Phys. Rev. C{\bf 69}, 034611 (2004).
 \bibitem{Ben03} 
 M. Bender, P.-H. Heenen, and P.-G. Reinhard,
  Rev. Mod. Phys.  {\bf 75}, 121 (2003).
\bibitem{Vre05} 
D. Vretenar, A.V. Afanasjev, G.A. Lalazissis, and P. Ring,
Phys. Rep. {\bf 409}, 101 (2005).
\bibitem{Peru08} 
S. P\'eru and H. Goutte,
Phys. Rev. {\bf C77}, 044313 (2008).
\bibitem{Lo10} 
C. Losa, A. Pastore, T. Dossing, E. Vigezzi, and R.A. Broglia,
Phys. Rev. {\bf C81}, 064307 (2010).
\bibitem{Gupta_15} 
Y.K. Gupta, U. Garg, J.T. Matta, D. Patel, T. Peach et al,
Phys. Lett. {\bf B748}, 343 (2015).
\bibitem{Kva15} 
J. Kvasil, V.O. Nesterenko, A. Repko, P.-G. Reinhard, and W. Kleinig,
EPJ Web of Conf. {\bf 107}, 05003 (2016).
\bibitem{Yosh10} 
  K. Yoshida,
Phys. Rev.{\bf C82}, 034324 (2010).
\bibitem{Kva_Is} 
 J. Kvasil, V.O. Nesterenko, A. Repko, D. Bozik, W. Kleinig and P.-G. Reinhard,
J. Phys.: Conf. Series {\bf 580}, 012053 (2015).
\bibitem{Yosh13} 
  K. Yoshida and T. Nakatsukasa,
Phys. Rev.{\bf C88}, 034309 (2013).
\bibitem{Peru11} 
S. P\'eru, G. Gosselin, M. Martini, M. Dupuis, S. Hilaire, and J.-C. Devaux,
Phys. Rev. {\bf C83}, 014314 (2011).
\bibitem{Kva_PS} 
J. Kvasil, D. Bo\v{z}ik, A. Repko, P.-G. Reinhard, V.O. Nesterenko, and W. Kleinig,
Phys. Scr. {\bf 90}, 114007 (2015).
\bibitem{bnl_exp} 
Evaluated Nuclear Structure Data File [http://www.nndc.bnl.gov].
\bibitem{Ri80} 
  P. Ring and P. Schuck, {\it Nuclear Many Body Problem},
  (Springer-Verlag, New York, 1980).
\bibitem{Repko} A. Repko, J. Kvasil, V.O. Nesterenko, and P.-G. Reinhard,
 arXiv:1510.01248[nucl-th].
\bibitem{Ne02} 
 V.O. Nesterenko, J. Kvasil, and P.-G. Reinhard,
 Phys. Rev. {\bf C66}, 044307 (2002).
\bibitem{Ne06} 
V.O.  Nesterenko, W. Kleinig, J. Kvasil, P. Vesely, P.-G. Reinhard, and D.S. Dolci,
 Phys. Rev. {\bf C74}, 064306 (2006).
\bibitem{Erl14a}
J. Erler and P.-G. Reinhard, J. Phys. {\bf G42}, 034026 (2014).
\bibitem{Rei15c}
P.-G. Reinhard, Phys. Scr. {\bf 91}, 023002 (2015).
\bibitem{Ben00} 
 M. Bender, K. Rutz, P.-G. Reinhard, and J.A. Maruhn,
 Eur. Phys. J. A{\bf 8}, 59 (2000).
 \bibitem{Kl09} 
  P. Klupfel, P.-G. Reinhard, T.J. Burvenich, and J.A. Maruhn,
  Phys. Rev. {\bf C79}, 034310 (2009).
\bibitem{SKPd} 
J. Dobaczewski, W. Nazarewicz, T. R. Werner, J.-F. Berger,
C. R. Chinn, and J. Decharg\'e, Phys. Rev. {\bf C53}, 2809 (1996).
\bibitem{Raman}
  S. Raman, C.W. Nestor, Jr., and P. Tikkanen,
  At. Data and Nucl. Data Tables {\bf 78}, 1 (2001).
\bibitem{Scamps_PRC_14}
 G. Scamps and D. Lacroix,
 Phys. Rev. C{\bf 89}, 034314 (2014).
 \bibitem{Cast}
 R.F. Casten, {\it Nuclear Structure from a Simple Perspective}
(Oxford Univercity Press, 1990).
\end{thebibliography}
\end{document}